\documentclass{svmult}

\usepackage{mathptmx}       
\usepackage{helvet}         
\usepackage{courier}        
\usepackage{type1cm}        
\usepackage{makeidx}         
\usepackage{graphicx}        
\usepackage{multicol}        
\usepackage[bottom]{footmisc}

\usepackage{color}        	

\makeindex             

\usepackage{amssymb,latexsym,amsmath}
\usepackage[numeric,initials,bibtex-style,nobysame]{amsrefs}

\usepackage{eucal}
\usepackage{mathrsfs}

\newcommand{\DD}{\ensuremath{\mathcal{D}}}
\newcommand{\G}{\ensuremath{\mathcal{G}}}
\renewcommand{\S}{\mathscr{S}}

\newcommand{\Ginf}{\ensuremath{\G^\infty}}

\newcommand{\mb}[1]{\ensuremath{\mathbb{#1}}}

\newcommand{\R}{\mb{R}}
\newcommand{\C}{\mb{C}}

\renewcommand{\d}{\ensuremath{\partial}}

\newfont{\bl}{msbm10 scaled \magstep2}
\newcommand{\td}{\mathrm{d}}
\newcommand{\tD}{\mathrm{D}}

\newcommand{\beq}{\begin{equation}}
\newcommand{\eeq}{\end{equation}}

\newcommand{\col}{\colon}

\newcommand{\F}{\ensuremath{{\mathcal F}}}

\newcommand{\dis}[2]{\langle #1 , #2 \rangle}
\newcommand{\inp}[2]{\langle #1 | #2 \rangle}  
\newcommand{\notmid}{\mid\kern-0.5em\not\kern0.5em}

\newcommand{\norm}[2]{{\| #1 \|}_{#2}}

\newcommand{\al}{\alpha}
\newcommand{\be}{\beta}
\newcommand{\ga}{\gamma}

\newcommand{\de}{\delta}
\newcommand{\eps}{\varepsilon}

\newcommand{\vphi}{\varphi}

\newcommand{\Om}{\Omega}
\newcommand{\sig}{\sigma}

\newcommand{\supp}{\mathop{\mathrm{supp}}}

\newcommand{\ovl}[1]{\overline{#1}}

\begin{document}

\title*{Generalized solutions and distributional shadows for Dirac equations}
\author{G\"unther H\"ormann and Christian Spreitzer}
\institute{G\"unther H\"ormann \at Fakult\"at f\"ur Mathematik, Universit\"at Wien, Austria, \email{guenther.hoermann@univie.ac.at}
\and Christian Spreitzer \at P\"adagogische Hochschule Nieder\"osterreich, Austria, \email{christian.spreitzer@ph-noe.ac.at}}
%
%
\maketitle


\abstract{We discuss the application of recent results on generalized solutions to the Cauchy problem for hyperbolic systems to Dirac equations with external fields. In further analysis we focus on the question of existence of associated distributional limits and derive their explicit form in case of free Dirac fields with regularizations of initial values corresponding to point-like  probability densities.}

\section{Introduction}

The Dirac equation on Minkowski space $M$ describes a relativitic spin-$\frac{1}{2}$ particle field $\psi \col M \to \C^4$,
involving the Dirac $4 \times 4$-matrices $\ga^\alpha$ ($\alpha = 0,\ldots,3$) as generators of (a representation of) the Clifford algebra. Two prominent examples of Dirac equations with external fields arise in the following models (cf.\ \cite{Thaller:92}):
\begin{trivlist}
\item 1. If an electromagnetic potential one-form $A$ (with components $A_\alpha$) is given on $M$, then the action on a particle with charge $e$ and mass $m$ is described by
\beq\label{Dirac1}
     (\d_t - i e A_0) \psi + \sum_{j=1}^3 \ga^0 \ga^j (\d_j  - i e A_j) \psi + i m \ga^0 \psi 
     = 0.
\eeq

The distribution theoretic Cauchy problem with compactly supported smooth $A$ and initial data given on arbitrary Cauchy surfaces is reviewed in detail in \cite{DM:14}.

\item 2. If $F$ is a given electromagnetic field two-form (with components $F_{\alpha \beta}$), then the action on a neutral particle with magnetic moment $\mu$ and mass $m$ is described by
\beq\label{Dirac2}
  \d_t  \psi + \sum_{j=1}^3 \ga^0 \ga^j \d_j  \psi + \frac{\mu}{2} \sum_{\al, \be = 0}^3 \ga^0 \ga^\al \ga^\be F_{\al \be} \psi + i m \ga^0 \psi = 0.
\eeq
\end{trivlist}

Both of the above cases will be seen to be combined in Section 2 below as special instances of a symmetric hyperbolic operator with generalized functions as coefficients. Moreover, this more general setting will enable us to approach the question of how to describe fields that are initially localized in a point-like sense. Recall that for a solution $\psi$ of the Dirac equation, the quantity $|\psi(t,.)|^2$ is interpreted as a \emph{spatial probability density at time t.}  What could we then say about an initially prescribed probability density $|\psi(0,.)|^2$ modeling a $\de$-type concentration? 

In terms of a representative $(\psi_\eps)_{\eps > 0}$ of a generalized spinor field, this would mean that we solve the Dirac equation with an initial condition represented in the form $\psi_\eps {\mid}_{t = 0} = \vphi_\eps$, where the regularization family $(\vphi_{\varepsilon})_{\eps > 0}$ has the property that $|\varphi_{\varepsilon}|^2$ converges to $\de$ in the sense of distributions as $\eps \to 0$. In this sense,  $(\vphi_{\varepsilon})_{\eps > 0}$ is a regularization of $\sqrt{\delta}$. The corresponding Cauchy problem is well-posed in spaces of Colombeau generalized functions according to the results presented in Section 2. Moreover, in case of the free Dirac equation we will prove in Section 3 that the probability density associated with the generalized solution possesses a distributional limit and we will determine its explicit shape. Related questions for the Schr\"odinger equation have been discussed in \cite{Hoermann:11,Hoermann:17}, where also some details on choices of regularizations of $\sqrt{\delta}$ can be found.

Why is it important to allow for a non-smooth potential $A$ and a non-smooth electromagnetic field $F$? Recall that the Maxwell equations in terms of exterior differentials are $\td F = 0$, $\ast \td \!\ast F = J$, where $J$ is the current one-form on $M$ and $\ast$ denotes the Hodge-star operator according to the Lorentz metric $g$ on $M$.
 For the electromagnetic field generated by a particle of charge $e$ and with world line $s \mapsto c(s)$, we obtain the current $J$ as the distributional one-form on $M$ supported on $c$, acting on a compactly supported vector field $v$ by 
$$
    \dis{J}{v} = e \int_{-\infty}^\infty g\big(\dot c(s), v(c(s))\big)  \, \td s.
$$   

\begin{remark} (i) 
The distributional nature of $J$ (and hence $F$) is well-known to cause serious mathematical problems with self-interaction due to the non-smooth \emph{Lorentz force} $L$ generated in the form
$\dis{L}{v} = e \int_{-\infty}^\infty F_{(c(s))} \big(\dot c(s),v(c(s))\big)\, \td s$. Coupling with the Maxwell equations should include the \emph{radiation-reaction} of the electron, but leads to systems of nonlinear differential equations with distributional data. A rigorous derivation of the Lorentz-Dirac equation in this context can be found in \cite{Gsponer:08a}. The generalized solutions to the coupled field equations and the equation of motion have been analyzed in \cite{HK:98,HK:00}, where also the non-existence of distributional shadows has been shown. These difficulties do not disappear upon quantization, but some renormalization effects could be described via generalized numbers  in \cite{GC:08, Gsponer:08b}. Discussing instead Dirac equations with external fields might thus be excused to some extent by the following folklore wisdom (e.g., from \cite{Thaller:92}): `A theory of particles in an ``external'' field is a first step towards a description of a true interaction.'

\noindent (ii) 
In terms of geometric structures, it is natural to consider the electromagnetic field as a curvature two-form on a space-time $(M,g)$. The electromagnetic field $F$ on $M$ then stems from a connection on a principal $U(1)$-bundle $P$ over $M$ which is given in terms of a one-form $\omega$ on $P$. If $M$ is contractible (e.g.\ Minkowski space), then $P = M \times U(1)$ is trivial and we have the simple description 
$\omega = \frac{\td z}{z} - i A$, where $z$ denotes the coordinate on $\C \supset U(1)$ and  $A$ is a real one-form on $M$. If $V$ is a complex vector space (with $U(1)$ action) and a field $\psi$ is (locally) written as a map from $M$ into $V$, then the covariant derivative in the direction of the tangent vector field $h$ is $\tD_h \psi = \td\psi (h) - i A(h) \psi$. We obtain the curvature two-form $R(h_1,h_2)(\psi) := \tD_{h_1} (\tD_{h_2} \psi ) - \tD_{h_2} (\tD_{h_1} \psi) - \tD_{[h_1,h_2]} \psi 
     = - i \td A(h_1,h_2) \psi $, which means $F = iR = \td A$. 
All these geometric constructions  can be carried out with non-smooth fields in the sense of generalized connections and curvature as developed in \cite{KSV:05}, where also an application to Dirac's theory of magnetic monopoles and basics of a Yang-Mills theory are given. A natural next step along these lines would be to discuss generalized spinor bundles and Dirac operators in this sense as well.
\end{remark}

The contents of the remaining parts is as follows: Section 2 presents a brief review of the Cauchy problem and applications to Dirac equations with non-smooth external fields. Section 3 discusses distributional shadows of free Dirac fields with $\sqrt{\delta}$ as initial data in $1+1$ dimensions and on Minkowski space.

\section{The Cauchy problem for the Dirac equation with external fields on Minkowski space}

All the results described in this section can be found in \cite{Spreitzer:17} or follow directly from statements proved there.
Let $T > 0$ and $\Om_T := [0,T] \times \R^3$. The two model equations \eqref{Dirac1} and \eqref{Dirac2} may be combined in the form of the following first-order symmetric hyperbolic system for a Colombeau generalized spinor field $\psi \in \G_{L^2}(\Om_T)^4$
\beq\label{DiracGeneral}
  \d_t  \psi + \sum_{j=1}^3 \ga^0 \ga^j \d_j  \psi + B  \psi = 0, \quad \psi{\mid}_{t = 0} = \vphi,
\eeq
where $B$ is a matrix with components $b_{kl} \in \G_{L^\infty}(\Om_T)$, and initial value $\vphi \in \G_{L^2}(\R^3)^4$.

\begin{theorem} If $B + B^*$ is of $L^{1,\infty}$-log-type, that is $\int_0^T\norm{(b_{kl,\eps} + \ovl{b_{lk,\eps}})(t,.)}{L^\infty}\, \td t = {\mathcal O}(\log(\frac{1}{\eps}))$ as $\eps \to 0$, then  there exists a unique solution $\psi \in \G_{L^2}(\Om_T)^4$ to \eqref{DiracGeneral}.
\end{theorem}

We point out that the theorem applies with $B$ containing the field of a moving point particle, if logarithmic scaling is used in its regularization. We note in passing that there is also an intrinsic regularity property  phrased in terms of $\Ginf$, which requires uniform asymptotic bounds on all derivatives (\cite{HS:12}): If $\vphi\in\Ginf(\mathbb R^3)^4$ and $B$ is of logarithmic slow scale in the $\Ginf$-sense, then the solution satisfies  $\psi \in \Ginf(\Om_T)^4$.

Regarding distributional aspects and compatibility with classical  spaces, we have the following results. Recall that $u \in \DD'$ is said to be a distributional shadow of $\psi \in \G$, in notation $\psi \approx u$, if $\psi_\eps \to u$ ($\eps \to 0$) holds in $\DD'$ for any (hence every) representative.

\begin{proposition} (i) If $\vphi$ and $B$ are smooth, then $\psi$ equals the unique smooth solution.
 
\noindent (ii) If $\vphi$ has components in $H^s$ and $B$ is smooth, then $\psi \approx v$, where $v$ denotes the unique distributional solution.

\noindent (iii) If $\vphi\in H^1(\R^3)^4$ and $B$ has components in $L^1([0,T],W^{1,\infty}(\R^3))$, then $\psi \approx u$ with  $u\in C([0,T],H^1(\R^3))^4 \cap W^{1,1}(\Om_T)^4$.

\noindent (iv) If $\vphi \in L^2(\R^3)^4$ and $B$ has components in $L^1([0,T],H^2(\R^3))$, then   $\psi \approx u$ with $u \in C([0,T],L^2(\R^3))^4$. 

\end{proposition}

Unfortunately, none of these convergence results are applicable to the situation of a coefficient matrix $B$ involving the field of a moving point particle or an initial value $\vphi$ corresponding to a point-like concentrated field configuration.

\section{Distributional limits for free Dirac fields with $\sqrt{\delta}$ initial data}

\subsection{The case of one spatial dimension}

We consider the Cauchy problem for a free Dirac particle of mass $m$ in 1+1 dimensions with generalized functions as initial data (and physical units with $\hbar =c=1$),
\begin{eqnarray*}
	i\partial_t \psi + i\sigma^1 \partial_x\psi- m \,\sigma^3 \psi
	&=& 0 \\
	\psi|_{t=0}
	&=&	 \varphi  
\end{eqnarray*}
where  $\varphi$ is represented by $(\varphi_{\varepsilon})_{\varepsilon >0}$, $\sigma^1=\left(\begin{smallmatrix}  0 & 1 \\ 1 & 0\end{smallmatrix}  \right)$,  $\sigma^3=\left(\begin{smallmatrix}  1 & 0 \\ 0 & -1 \end{smallmatrix}  \right)$.  
We are interested in the distributional limit of $|\psi_\eps|^2$, where $(\psi_\eps)_{\eps > 0}$ represents the generalized solution, 
if $|\varphi_\eps|^2$ converges to $\delta$. For rapidly decaying initial data we may write 
\begin{multline*}
\!\!\!\!\!\!\!\!
\psi_{\varepsilon}(t,x) =\int\limits_{\mathbb R}\!\!\frac{e^{ikx}}{\!\!\sqrt{2\pi}} \Big( \langle u_{\text{pos}}(k)|\widehat{\varphi_{\eps}}(k) \rangle
  u_{\text{pos}}(k) e^{-it\lambda(k)} + 
  \langle u_{\text{neg}}(k)|\widehat{\varphi_{\eps}}(k) \rangle u_{\text{neg}}(k) e^{it\lambda(k)}\Big)  \td k
\end{multline*}
where $\widehat{\varphi}$ denotes the Fourier transform of $\varphi$ and $\lambda(k):=\sqrt{k^2+m^2}$. The functions  
$$ 
u_{\text{pos}}(k)=\frac{1}{\sqrt 2} \left(\begin{array}{c} \sqrt{1+\frac{m}{\lambda(k)} } \\  \mathrm{sgn}(k)\sqrt{1-\frac{m}{\lambda(k)}}  \end{array}\right),\qquad u_{\text{neg}}(k)=\frac{1}{\sqrt 2}\left(\begin{array}{c} -\mathrm{sgn}(k)\sqrt{1-\frac{m}{\lambda(k)}} \\ \sqrt{1+\frac{m}{\lambda(k)} }  \end{array}\right)
$$
  are normalized eigenvectors of the matrix  $-ik\sigma^1+ m \,\sigma^3 $ with eigenvalues $\pm\lambda(k)$, corresponding to positive and negative energies.   To model the initial wave-function in $\mathcal G_{L^2}(\mathbb R)^2$, we pick  $\rho_1,\rho_2\in \mathcal S(\mathbb R)$ such that $\norm{\rho_1}{L^2(\mathbb R)}^2+\norm{\rho_2}{L^2(\mathbb R)}^2=1$ and set $\varphi(x):=\left(\begin{smallmatrix} \rho_1(x) \\ \rho_2(x) \end{smallmatrix}\right)$. The scaling  $\varphi_{\eps}(x):=\frac{1}{\sqrt{\eps}}\varphi(\frac{x}{\eps})$ yields $|\varphi_{\eps}|^2\to \delta$ as $ \eps\to 0$. Interpreting $\mu_t^{\eps}(x):=|\psi_\eps(t,x)|^2$ as a spatial probability density at time $t$, we consider its distributional action on a test function  $h\in \mathcal D(\mathbb R)$ and use Fubini's theorem and the fact that    $\widehat{\varphi_{\eps}}(k)=\sqrt{\eps}\widehat{\varphi}(\eps k)$ to obtain
\begin{multline*}
 \langle \mu_t^{\eps},h\rangle=
 \frac{\varepsilon}{2\pi}\int\limits_{\mathbb R}\int\limits_{\mathbb R}\int\limits_{\mathbb R}e^{i(k'-k)x}h(x)\td x\\
 \bigg(e^{-it(\lambda(k')-\lambda(k))}\overline{\langle u_{\text{pos}}(k)|\widehat{\varphi}(\eps k)\rangle} \langle u_{\text{pos}}(k')|\widehat{\varphi}(\eps k')\rangle  \langle u_{\text{pos}}(k)|u_{\text{pos}}(k')\rangle \\
 + 2\text{Re}\Big(e^{it(\lambda(k')+\lambda(k))}\overline{\langle u_{\text{pos}}(k)|\widehat{\varphi}(\eps k)\rangle} \langle u_{\text{neg}}(k')|\widehat{\varphi}(\eps k')\rangle  \langle u_{\text{pos}}(k)|u_{\text{neg}}(k')\rangle\Big) \\
  + e^{it(\lambda(k')-\lambda(k))}\overline{\langle u_{\text{neg}}(k)|\widehat{\varphi}(\eps k)\rangle} \langle u_{\text{neg}}(k')|\widehat{\varphi}(\eps k')\rangle  \langle u_{\text{neg}}(k)|u_{\text{neg}}(k')\rangle\bigg) \td k 
 \, \td k' .
 \end{multline*} 
By the change of  variables $(k,k')\mapsto (\eta,\xi):=(\eps k,k'-k)$ we may write  this as 
\begin{multline*}
 \langle \mu_t^{\eps},h\rangle=
 \frac{\varepsilon}{\sqrt{2\pi}}\int\limits_{\mathbb R}\int\limits_{\mathbb R}\mathcal F^{-1}(h)(\xi)\\
 \bigg(e^{-it\Lambda_{\eps}^{-}(\eta,\xi)}\overline{\langle u_{\text{pos}}({\textstyle \frac{\eta}{\eps}})|\widehat{\varphi}(\eta)\rangle} \langle u_{\text{pos}}({\textstyle\frac{\eta}{\eps}}+\xi)|\widehat{\varphi}(\eta+\eps\xi)\rangle  \langle u_{\text{pos}}({\textstyle \frac{\eta}{\eps}})|u_{\text{pos}}({\textstyle\frac{\eta}{\eps}}+\xi)\rangle \\
 + 2\text{Re}\Big(e^{it\Lambda_{\eps}^{+}(\eta,\xi)}\overline{\langle u_{\text{pos}}({\textstyle \frac{\eta}{\eps}})|\widehat{\varphi}(\eta)\rangle} \langle u_{\text{neg}}({\textstyle\frac{\eta}{\eps}}+\xi)|\widehat{\varphi}(\eta+\eps\xi)\rangle  \langle u_{\text{pos}}({\textstyle \frac{\eta}{\eps}})|u_{\text{neg}}({\textstyle\frac{\eta}{\eps}}+\xi)\rangle\Big) \\
  + e^{it\Lambda_{\eps}^{-}(\eta,\xi)}\overline{\langle u_{\text{neg}}({\textstyle \frac{\eta}{\eps}})|\widehat{\varphi}(\eta)\rangle} \langle u_{\text{neg}}({\textstyle\frac{\eta}{\eps}}+\xi)|\widehat{\varphi}(\eta+\eps\xi)\rangle  \langle u_{\text{neg}}({\textstyle \frac{\eta}{\eps}})|u_{\text{neg}}({\textstyle\frac{\eta}{\eps}}+\xi)\rangle\bigg) \td \eta 
 \, \td \xi 
 \end{multline*}
 where $\Lambda_{\eps}^{\mp}(\eta,\xi):=\lambda(\frac{\eta}{\eps}+\xi)\mp\lambda(\frac{\eta}{\eps})$. 
 The integrand is  bounded uniformly in $\eps$ by the integrable function $g(\eta,\xi):=\norm{\widehat{\varphi}}{L^{\infty}(\mathbb R)}|\widehat{\varphi}(\eta)| |\mathcal F^{-1}(h)({\xi})|$  and thus the limit $\eps\to 0$  commutes with integration by Lebesgue's dominated convergence.
Writing 
$ \lambda(k)=|k|(1+\frac{m^2}{k^2})^{1/2}=|k|+\mathcal O(|k|^{-1})$ as  $|k|\to \infty$
and noting that for all $\eta\neq 0$, 
$$\lim_{\eps \to 0}|{\textstyle \frac{\eta}{\eps}}+\xi|-|{\textstyle\frac{\eta}{\eps}}|=\mathrm{sgn}(\eta) \xi,$$
 it is easy to see that
$  \lim_{\eps\to 0} e^{\mp it(\lambda(\frac{\eta}{\eps}+\xi)-\lambda(\frac{\eta}{\eps}))}=  e^{\mp it\mathrm{sgn}(\eta)\xi }$.
Moreover we have
$u_{\text{pos}}({\textstyle \frac{\eta}{\eps}}) \to {\textstyle\frac{1}{\sqrt{2}}}\left(\begin{smallmatrix} 1 \\ \mathrm{sgn}(\eta)  \end{smallmatrix} \right)$ and  $u_{\text{neg}}({\textstyle \frac{\eta}{\eps}}) \to {\textstyle\frac{1}{\sqrt{2}}}\left(\begin{smallmatrix} -\mathrm{sgn}(\eta) \\ 1  \end{smallmatrix}\right)$  as $\eps \to 0$.
Using  these observations we can write the pointwise limit arranged in terms of $\widehat{\rho_1}$ and $\widehat{\rho_2}$, thereby obtaining 
\begin{multline*} 
 \lim\limits_{\eps\to 0} \langle \mu_t^{\eps},h\rangle = 
   \frac{1}{2\sqrt{2\pi}}\int\limits_{\mathbb R}\int\limits_{\mathbb R} \mathcal F^{-1}(h)({\xi}) \Big(e^{-it\text{sgn}(\eta)}
|\widehat{\rho_1}(k)+\mathrm{sgn}(\eta)\widehat{\rho_2}(k)|^2 \\
 + e^{it\text{sgn}(\eta)} |\widehat{\rho_1}(k)-\mathrm{sgn}(\eta)\widehat{\rho_2}(k)|^2 \Big)\td\eta \td\xi,
\end{multline*}
which upon splitting the integral according to the sign of $\eta$ and re-combining  yields
$$
  \lim\limits_{\eps\to 0} \langle \mu_t^{\eps},h\rangle ={\textstyle\frac{1}{2}}h(t)\norm{\rho_1+\rho_2}{L^2(\mathbb R)}^2  + {\textstyle\frac{1}{2}}h(-t)\norm{\rho_1-\rho_2}{L^2(\mathbb R)}^2,  
$$
which by the normalization  $\norm{\rho_1}{L^2}^2+\norm{\rho_2}{L^2}^2=1$ implies the distributional limit
$$ \lim\limits_{\eps \to 0} \mu_t^{\eps} =  \Big({\textstyle\frac{1}{2}}+\int\limits_{\mathbb R}\text{Re}(\overline\rho_1\rho_2)(k)\td k\Big) \delta_{t}
 +\Big({\textstyle\frac{1}{2}}-\int\limits_{\mathbb R}\text{Re}(\overline\rho_1\rho_2)(k)\td k\Big) \delta_{-t}.
$$
Hence the distributional wave-function of the particle at time $t$ is a  convex combination of $\delta_t$ and $\delta_{-t}$, where the coefficients are determined by $\mathrm{Re}\int (\overline{\rho_1}\rho_2)(k)\mathrm dk$. For example, if $\text{Re}(\overline\rho_1\rho_2)=0$, then $u_t^{\eps} \to \frac{1}{2}(\delta_t+\delta_{-t})$ as $\eps \to 0$. On the other hand,
if $\rho_1=\pm\rho_2$, then  $u_t^{\eps}\to \delta_{\pm t}$ as $\eps \to 0$.

\subsection{Minkowski space}

In physical units with $\hbar =c=1$, 
we may write the Dirac equation in the  `Hamiltonian operator form'  (cf.\ \cite[Chapter XIV]{DL:V5})
$$
   i \d_t \psi = -i \sum_{j=1}^3 \al^j  \d_{x_j} \psi + m \be \psi,
$$
where $\psi \in C(\R,\S'(\R^3))^4$,  and 
$\be = \left(\begin{smallmatrix} I_2 & 0\\ 0 & -I_2\end{smallmatrix}\right)$ and $\al^j = \left( \begin{smallmatrix} 0 & \sig^j\\ \sig^j & 0\end{smallmatrix}\right)$  ($j=1,2,3$)
with $I_2$ denoting the $(2 \times 2)$-identity matrix and with the Pauli matrices
$\sig^1 = \left(\begin{smallmatrix} 0 & 1\\ 1 & 0\end{smallmatrix}\right)$, 
   $\sig^2 = \left(\begin{smallmatrix} 0 & -i\\ i & 0\end{smallmatrix}\right)$, 
   $ \sig^3 = \left(\begin{smallmatrix} 1 & 0\\ 0 & -1\end{smallmatrix}\right)$.
Writing $u = \exp(i m \be t) \psi$ we obtain
$\d_t u + A(\d_x) u = 0$ with $A(\d_x) := \sum_{j=1}^3 \al^j \d_{x_j}$.
Given an initial condition 
$\psi \mid_{t=0} = \vphi \in \S'(\R^3)^4$
we obtain the unique solution in the form (cf.\ \cite[Chapter XIV]{DL:V5})
$$
    \exp(i m \be t) \psi = u(t) = E_0(t) \ast \vphi - E_1(t) \ast A(\d_x) \vphi,
$$
where the convolution is spatial componentwise and
$E_0 (t) := \F^{-1}(\cos(t |.|))$, $E_1(t) := \F^{-1}\left(\frac{\sin(t |.|)}{|.|}\right)$.
If $\vphi(t) \in L^2(\R^3)^4$ then $\norm{\psi(t)}{L^2} = \norm{u(t)}{L^2} = \norm{\vphi}{L^2}$ for every $t$ by unitarity of the time evolution in $L^2(\R^3)$ and also by unitarity of $\exp(i m \be t)$ on $\C^4$ which implies $|\psi(t,x)|^2 = |u(t,x)|^2$.
 
Let $\psi_\eps$ denote the unique solution with initial values   $\vphi_\eps(x):= \eps^{-3/2}\vphi(x/\eps)$, where $\vphi\in \S(\R^3)^4$ and $\norm{\vphi}{L^2}=1$.  We have $|\vphi_\eps|^2 \to \delta$ in $\S'(\R^3)$ as $\eps \to 0$.

Let $\mu_t^\eps(x) := |\psi_\eps(t,x)|^2$ denote the spatial probability density at time $t$ corresponding to the solution $\psi_\eps$. By unitarity of the time evolution, we have in terms of finite measures
$\dis{\mu_t^\eps}{1} = \norm{\psi_\eps(t,.)}{L^2}^2 = 
    \norm{\vphi_\eps}{L^2}^2 = \norm{\vphi}{L^2}^2 = 1$, 
but in the sequel we are interested in the convergence properties of $\mu_t^\eps$ in $\S'(\R^3)$ as $\eps \to 0$.

\begin{remark} (i) Note that with the initial value $\vphi_\eps$ specified above, a simple calculation yields $\vphi_\eps \to 0$ in $\S'(\R^3)^4$ and therefore $\psi_\eps(t,.) \to 0$ in $\S'(\R^3)^4$ by continuity of the solution map.

\noindent (ii) If instead we consider initial values $\vphi_\eps(x) = \eps^{-3}\vphi(x/\eps)$ as regularizations of delta in the sense that $|\vphi_{\eps}|\to \delta$ in $\S'$ as $\eps \to 0$
(recall that $\vphi(x)\in \mathbb C^4$  and that $\norm{\vphi}{L^2}=1$), 
e.g.,  with $\supp(\vphi)$ compact, and denote again by $\psi_\eps$ the corresponding unique solution, then $|\psi_\eps(t,.)|^2$ diverges in $\S'(\R^3)$ for every $t$ as $\eps \to 0$: We have finite speed of propagation for the Dirac equation, hence $\supp(\psi_\eps(t,.)) \subseteq \supp(\vphi) + B_{t}(0) =: K_t$ and picking $h \in \DD(\R^3)$ such that $h \geq 0$ and $h = 1$ on $K_t$ we deduce
$\dis{|\psi_\eps(t,.)|^2}{h} \geq \norm{\psi_\eps(t,.)}{L^2}^2 = 
   \norm{\vphi_\eps}{L^2}^2 = 
   \frac{\norm{\vphi}{L^2}^2}{\eps^3} \to \infty$ as  $\eps \to 0$. 
\end{remark}

Addressing now the question of convergence of $\mu_t^\eps$ at fixed $t$ as $\eps \to 0$, we will represent the action of $\mu_t^\eps$ as a distribution on a test function $h \in \DD(\R^3)$ by employing the decomposition of the solution $\psi_{\eps}$ in  terms of  eigenfunctions corresponding to positive or negative energy respectively helicity  as in \cite[Chapter 1, Appendix 1.F]{Thaller:92}:
Introducing      
 $a_{\pm}(k):=\frac{1}{\sqrt 2}\sqrt{1\pm \frac{m}{\lambda(k)}}$, where $\lambda(k)=(k^2+m^2)^{1/2}$, 
    $h_+(k):=\frac{1}{\sqrt{2|k|(|k|-k_3)}}\left(\begin{smallmatrix} k_1-ik_2 \\ |k|-k_3  \end{smallmatrix} \right)$, and $h_-(k):=\frac{1}{\sqrt{2|k|(|k|-k_3)}}\left(\begin{smallmatrix} k_3-|k| \\ k_1+ik_2  \end{smallmatrix} \right)$, we define
$$
\omega_{\text{pos},\pm}(k):= \left( \begin{array}{c} a_+(k)h_{\pm}(k)\\\pm a_-(k)h_{\pm}(k)\end{array}\right), \qquad  \omega_{\text{neg},\pm}(k):= \left( \begin{array}{c} \mp a_-(k)h_{\pm}(k)\\ a_+(k)h_{\pm}(k)\end{array}\right).
$$ 
Note that
 $h_{\pm}$ is homogeneous of degree 0. Moreover, $\lim_{|k|\to \infty} a_{\pm}(k)=1/\sqrt 2$ and thus  $\lim_{\eps\to 0}\omega_{\text{pos},\pm}({\textstyle  k /\eps})$, $\lim_{\eps\to 0}\omega_{\text{neg},\pm}({\textstyle k/\eps})$ are  homogeneous of degree 0 as well.
The four complex-valued vectors  $\omega_{\text{pos},\pm}(k)$, $\omega_{\text{neg},\pm}(k)$ form an orthonormal system with respect to the inner product $\langle\:\:|\:\:\rangle_{\mathbb C^4}$. For initial data $\varphi_\eps\in L^2(\mathbb R^3)$, we may write 
\begin{multline*} 
\psi_\eps(t,x)=\\ (2\pi)^{-3/2}\int_{\mathbb R^3}\bigg( e^{i\langle k|x\rangle - i\lambda(k)t}\Big( \langle \omega_{\text{pos}}^+(k)|\widehat{\varphi_\eps}(k)\rangle\omega_{\text{pos}}^+(k)+ \langle \omega_{\text{pos}}^-(k)|\widehat{\varphi_\eps}(k)\rangle\omega_{\text{pos}}^-(k)\Big)
\\+ e^{i\langle k|x\rangle + i\lambda(k)t}\Big( \langle \omega_{\text{neg}}^+(k)|\widehat{\varphi_\eps}(k)\rangle\omega_{\text{neg}}^+(k)+ \langle \omega_{\text{neg}}^-(k)|\widehat{\varphi_\eps}(k)\rangle\omega_{\text{neg}}^-(k)\Big)\bigg) \td k
\end{multline*}
and obtain 
\begin{multline*} 
(2\pi)^{3}\dis{\mu_t^\eps}{h}=(2\pi)^{3}\langle |\psi_{\eps}(t,\cdot)|^2,h \rangle =  \int\limits_{\mathbb R^3\times \mathbb R^3\times \mathbb R^3} \Big(e^{i\langle k-k'|x\rangle}   \\
e^{it\left(\lambda(k)-\lambda(k')\right)}\big(
\langle \omega_{\text{pos}}^+(k) |\widehat{\varphi_{\eps}}(k)\rangle \overline{\langle \omega_{\text{pos}}^+(k')|\widehat{\varphi_{\eps}}(k')\rangle}  \langle \omega_{\text{pos}}^+(k') |\omega_{\text{pos}}^+(k) \rangle \\
+
\langle \omega_{\text{pos}}^-(k) |\widehat{\varphi_{\eps}}(k)\rangle \overline{\langle \omega_{\text{pos}}^-(k')|\widehat{\varphi_{\eps}}(k')\rangle}  \langle \omega_{\text{pos}}^-(k') |  \omega_{\text{pos}}^-(k) \rangle 
\big) +\\
e^{-it\left(\lambda(k)-\lambda(k')\right)} 
\big(
\langle \omega_{\text{neg}}^+(k) |\widehat{\varphi_{\eps}}(k)\rangle \overline{ \langle \omega_{\text{neg}}^+(k')|\widehat{\varphi_{\eps}}(k')\rangle} \langle \omega_{\text{neg}}^+(k') |\omega_{\text{neg}}^+(k) \rangle \\
+
\langle \omega_{\text{neg}}^-(k) |\widehat{\varphi_{\eps}}(k)\rangle \overline{ \langle \omega_{\text{neg}}^-(k')|\widehat{\varphi_{\eps}}(k')\rangle}  \langle \omega_{\text{neg}}^-(k') |  \omega_{\text{neg}}^-(k) \rangle  \big) +\mathcal N_{\eps}(k,k')\Big) h(x)\td (k,k',x), 
\end{multline*}
where the symbol $\mathcal N_{\eps}(k,k')$ represents all terms involving `mixed'  products  such as $\langle \omega_{\text{pos}}^{+}(k') |  \omega_{\text{neg}}^{+}(k) \rangle$ or  $\langle \omega_{\text{pos}}^{+}(k') |  \omega_{\text{pos}}^{-}(k) \rangle$.   Using that  $\widehat{\varphi_{\eps}}(k)=    \eps^{3/2}\widehat{\varphi}(\eps k)$ and changing variables according to $k\mapsto \frac{\eta}{\eps}$ and $k'\mapsto \frac{\eta}{\eps}+\xi$, we find
\begin{multline*} 
(2\pi)^3\lim\limits_{\eps\to 0}\dis{\mu_t^\eps}{h} =  \lim\limits_{\eps\to 0}\int_{\mathbb R^3\times \mathbb R^3\times \mathbb R^3}\Bigg(  \mathcal N_{\eps}({\textstyle \frac{\eta}{\eps}},{\textstyle \frac{\eta}{\eps}+\xi})+ e^{-i\langle \xi|x\rangle-it\left(\lambda(\frac{\eta}{\eps}+\xi)-\lambda(\frac{\eta}{\eps})\right)} \\ 
\bigg(
\langle \omega_{\text{pos}}^+({\textstyle\frac{\eta}{\eps}}) |\widehat{\varphi}(\eta)\rangle\overline{\langle \omega_{\text{pos}}^+({\textstyle\frac{\eta}{\eps}}+\xi)|\widehat{\varphi}(\eta+\eps\xi)\rangle}  \langle \omega_{\text{pos}}^+({\textstyle\frac{\eta}{\eps}}+\xi) |\omega_{\text{pos}}^+({\textstyle\frac{\eta}{\eps}}) \rangle \\
+
\langle \omega_{\text{pos}}^-({\textstyle\frac{\eta}{\eps}}) |\widehat{\varphi}(\eta)\rangle\overline{\langle \omega_{\text{pos}}^-({\textstyle\frac{\eta}{\eps}}+\xi)|\widehat{\varphi}(\eta+\eps\xi)\rangle}  \langle \omega_{\text{pos}}^-({\textstyle\frac{\eta}{\eps}}+\xi) |  \omega_{\text{pos}}^-({\textstyle\frac{\eta}{\eps}}) \rangle  
\bigg) \Bigg)h(x)\td (\eta,\xi,x)\\
+
\lim\limits_{\eps\to 0}\int_{\mathbb R^3\times \mathbb R^3\times \mathbb R^3} e^{-i\langle \xi|x\rangle+it\left(\lambda(\frac{\eta}{\eps}+\xi)-\lambda(\frac{\eta}{\eps})\right)} \\
\bigg(
\langle \omega_{\text{neg}}^+({\textstyle\frac{\eta}{\eps}}) |\widehat{\varphi}(\eta)\rangle\overline{\langle \omega_{\text{neg}}^+({\textstyle\frac{\eta}{\eps}}+\xi)|\widehat{\varphi}(\eta+\eps\xi)\rangle}  \langle \omega_{\text{neg}}^+({\textstyle\frac{\eta}{\eps}}+\xi) |\omega_{\text{neg}}^+({\textstyle\frac{\eta}{\eps}}) \rangle \\
+
\langle \omega_{\text{neg}}^-({\textstyle\frac{\eta}{\eps}}) |\widehat{\varphi}(\eta)\rangle\overline{\langle \omega_{\text{neg}}^-({\textstyle\frac{\eta}{\eps}}+\xi)|\widehat{\varphi}(\eta+\eps\xi)\rangle}  \langle \omega_{\text{neg}}^-({\textstyle\frac{\eta}{\eps}}+\xi) |  \omega_{\text{neg}}^-({\textstyle\frac{\eta}{\eps}}) \rangle  
\bigg) h(x) \td (\eta, \xi, x).
\end{multline*}
The integrand is uniformly dominated in $L^1$ and in the pointwise limit $\eps\to 0$, the inner products of  the eigenfunctions  simplify by orthonormality and homogeneity  (in particular, $\mathcal N_{\eps}({\textstyle \frac{\eta}{\eps}},{\textstyle \frac{\eta}{\eps}+\xi})\to 0$), and the exponential terms converge  thanks to 
\begin{multline*}
   \lambda({\textstyle\frac{\eta}{\eps}+\xi})- \lambda({\textstyle\frac{\eta}{\eps}})= |{\textstyle\frac{\eta}{\eps}+\xi}|-   |{\textstyle\frac{\eta}{\eps}}| +\mathcal O(\eps)  = 
      \frac{ | \frac{\eta}{\eps}+\xi|^2-|\frac{\eta}{\eps}|^2 }{|\frac{\eta}{\eps}+\xi| + |\frac{\eta}{\eps}| }+\mathcal O(\eps) \\
      = \frac{2 \inp{\eta}{\xi} + \eps |\xi|^2}{|\eta|+|\eta+\eps \xi| } +\mathcal O(\eps)\to
        \frac{2 \inp{\eta}{\xi}}{2 |\eta|} = \frac{\inp{\eta}{\xi}}{|\eta|},
\end{multline*}
which implies
 \begin{multline*} 
\lim\limits_{\eps\to 0}\dis{\mu_t^\eps}{h}=\\
(2\pi)^{-3/2}\int_{\mathbb R^3\times \mathbb R^3} e^{-it\frac{\langle \eta | \xi \rangle}{|\eta|}} 
\bigg(
|\langle \omega_{\text{pos}}^+(\eta) |\widehat{\varphi}(\eta)\rangle|^2    
+
|\langle \omega_{\text{pos}}^-(\eta) |\widehat{\varphi}(\eta)\rangle|^2 
\bigg) \widehat{h}(\xi)\td \eta \td \xi\\
+
(2\pi)^{-3/2}\int_{\mathbb R^3\times \mathbb R^3}  e^{it  \frac{\langle \eta | \xi \rangle}{|\eta|}} 
\bigg(
|\langle \omega_{\text{neg}}^+(\eta) |\widehat{\varphi}(\eta)\rangle|^2   
+
|\langle \omega_{\text{neg}}^-(\eta) |\widehat{\varphi}(\eta)\rangle|^2
\bigg) \widehat{h}(\xi)\td \eta \td \xi \\
= 
\int_{\mathbb R^3}
\Big(f_{\mathrm{pos}}(\eta) h(-t{\textstyle \frac{\eta}{|\eta|}}) 
+
f_{\mathrm{neg}}(\eta) h(t{\textstyle \frac{\eta}{|\eta|}})\Big)\td \eta 
=  \int_{\mathbb R^3} \big(f_{\mathrm{pos}}(-\eta)+ f_{\mathrm{neg}}(\eta)\big)h(t{\textstyle \frac{\eta}{|\eta|}})\td \eta,
\end{multline*}
where $f_{\mathrm{pos}/\mathrm{neg}}(\eta):=|\langle \omega_{\mathrm{pos}/\mathrm{neg}}^+(\eta) |\widehat{\varphi}(\eta)\rangle|^2    
+
|\langle \omega_{\mathrm{pos}/\mathrm{neg}}^-(\eta) |\widehat{\varphi}(\eta)\rangle|^2$. We finally obtain the limit as a distribution supported on the 2-sphere of radius $t$, namely 
$$
  \lim\limits_{\eps\to 0}\dis{\mu_t^\eps}{h}=  
  \int_{\mathbb S^2}f(\theta)  h(t\theta)  \td \theta, \qquad \text{where } 
  f(\theta):=\int_0^{\infty} r^2  \big(f_{\mathrm{pos}}(-r\theta)+ f_{\mathrm{neg}}(r\theta)\big)\td r.
$$

 \bibliography{gue}
\bibliographystyle{abbrv}


\end{document}